\documentclass[11pt,a4paper,english,notitlepage]{article}
\usepackage{mathpazo}
\usepackage[T1]{fontenc}
\usepackage[latin1]{inputenc}
\usepackage{fancyhdr}
\usepackage{babel}

\usepackage{graphicx}
\usepackage[authoryear]{natbib}
\usepackage[unicode=true, pdfusetitle,
 bookmarks=true,bookmarksnumbered=false,bookmarksopen=false,
 breaklinks=false,pdfborder={0 0 1},backref=false,colorlinks=false]
 {hyperref}

\pdfoutput=1



\makeatother

\begin{document}

\title{\textsc{The Roundtable - An Abstract Model of Conversation Dynamics}}

\date{\,}

\author{M. Mastrangeli%
\thanks{Distributed Intelligent Systems and Algorithms Laboratory (DISAL),
Ecole Politechnique Federale de Lausanne, Lausanne (CH)%
}, M. Schmidt%
\thanks{Department of Economics, University of Maryland, College Park (USA)%
} \ and L. Lacasa%
\thanks{Instituto de Fisica Interdisciplinar y Sistemas Complejos (IFISC,
CSIC-UIB), Palma de Mallorca (ES)%
}}
\maketitle
\begin{abstract}
Is it possible to abstract a formal mechanism originating schisms
and governing the size evolution of social conversations? In this
work a constructive solution to such problem is proposed: an abstract
model of a generic N-party turn-taking conversation. The model develops
from simple yet realistic assumptions derived from experimental evidence,
abstracts from conversation content and semantics while including
topological information, and is driven by stochastic dynamics. We
find that a single mechanism - namely the dynamics of conversational
party's individual fitness, as related to conversation size - controls
the development of the self-organized schisming phenomenon. Potential
generalizations of the model - including individual traits and preferences,
memory effects and more elaborated conversational topologies - may
find important applications also in other fields of research, where
dynamically-interacting and networked agents play a fundamental role. 
\end{abstract}
\textbf{Disclaimer}: \textit{all codes as well as their NetLogo
implementations utilized in and derived from this work are freely
available under request.}

\section{Introduction\label{sec:Introduction}}

Multi-party conversations are prime manifestations of collective socio-cultural
interactions. The Santa Fe Institute's Complex Systems Summer School
2009 was an excellent occasion for the authors to investigate this
statement from an alternative and formal vantage point. During all
meals, groups of up to 20 people comfortably clustered in quasi-circular
ensembles and entered into lively turn-taking conversations. An occasional
observer could easily discern that table conversations were not stable.
Not all participants seated around a specific table took part at all
times in a table-wide conversation. Usually, participants took part
in conversations that involved only a subset of the people seated
around that table. As a result, each table had multiple, separate
sub-conversations going on at the same time. Moreover, people taking
part in one of these parallel chats usually did not remain involved
in the same sub-conversation indefinitely, but tended to leave their
original sub-conversation and join another, possibly neighboring one
going on at their same table. Remarkably, all of this happened independently
of the very topics of conversation - that indeed fluctuated spatially
and temporally.

Was this set of behaviors accidental, or was there, on the contrary,
any general underlying mechanism driving the group size evolution
of conversations? This question has been addressed in social sciences
from several perspectives. The general issue of pointing out the sociological
factors that determine the group size of a conversation dates back
to the seminal works of Simmel \citet{Simmel}. The splitting up of
a single conversation into two or more sub-conversations, referred
to as \emph{schism} \citet{Sacks,Egbert,Goodwin-art,Goodwin-book,Parker,Schegloff,Goffman},
was investigated by Goffmann in a qualitative, context-sensitive interpretative
approach \citet{Goffman}, and by Sacks and collaborators in the context
of discourse analysis \citet{Sacks}. Several social features and
effects of schisming were also addressed, including schism-induced
turns \citet{Egbert}, multi-focused gatherings \citet{Goffman} and
co-occurrence of turn-taking systems \citet{Sacks,Goodwin-book}.
Specific behaviors were documented by concrete experiments, such as
video tape recordings \citet{Egbert,Goodwin-art}, everyday experience
\citet{Goffman}, or hypothesizing and reasoning methodology \citet{Simmel}.
Significantly less work addressed the analysis of conversations and
schisming processes from an abstract, context-free point of view.
Such an approach was probably difficult to pursue in earlier times,
as most data were empiric and therefore context-related. However,
in recent years it has been possible to circumvent these restrictions
thanks to the social simulation framework \citet{Byrne,miller-page,axelrod-evolution,axelrod-complexity,Axelrod-inbook,bandini}.
This was developed to improve the understanding of and to evaluate
strategies, explanatory and predictive schemes of the behavior of
social systems whenever - for \emph{e.g.} practical or ethical reasons
- gathering direct observations was impossible. Particularly, the
well-established \emph{agent-based model} (ABM) paradigm \citet{bonabeau,miller-page,bandini}
has proven fruitful to simulate complex collective phenomena in several
domains \citet{miller-page,Byrne,axelrod-complexity,Axelrod-inbook,schweitzer}.
Works on multi-party interactions were pursued in the context of artificial
intelligence, addressing specific challenges such as \emph{e.g.} human-robot
interaction \citet{bono} or pattern recognition \citet{Aoki}; interaction
structure in meetings, among which \emph{e.g.} co-present meetings
in smart meeting rooms for archival and assistive purposes \citet{Ishizaki,Stiefekhagen},
and remote interaction \citet{vertegaal}; and in more general scenarios
\citet{schweitzer,miller-page}. Closer to our interests, Galley \emph{et
al.} proposed a remarkable topic segmentation algorithm for multi-party
speech, which is domain-independent but nonetheless content- and knowledge-sensitive
\citet{galley}. To our knowledge, no work so far addressed turn-taking
conversation dynamics from a purely-formal perspective, abstracting
from both content and context.

In this paper we introduce such an approach by describing the emergence
of conversational schisming as a self-organized complex collective
phenomenon. We present an abstract model, based on simple formal yet
realistic rules and driven by stochastic dynamics, that predicts the
time-evolving size of conversation groups. By embodying the rules
and boundary conditions into an ABM, we analyze how schisming drives
the development of conversations. Since our model is deliberately
abstract and context-free, our conclusions are general and do not
restrict to any particular class of turn-taking conversations. Furthermore,
proposed generalizations of our model may find useful applications
in other research areas, as well.\medskip{}

The rest of the paper is organized as follows: in Section 2 we present
the baseline conversational model, defining the basic agent rules.
In Section 3 we implement such model and provide the results of several
simulation runs; particularly, we distinguish between transient dynamics,
like real-life conversations, from stationary dynamics, which is the
(probably unrealistic) asymptotic limit of the dynamics which, nonetheless,
can in some cases be solved analytically in a mean-field approximation.
Finally, we provide a summary of conclusions, as well as a discussion
on the possible model generalizations and further applications of
the work.

\section[Assumptions]{The baseline model: assumptions\label{sec:Assumptions}}

We define our baseline conversation model by instantiating a set of
simplifying yet realistic assumptions: 
\begin{enumerate}
\item \emph{Homogeneous initial conditions}. At the beginning, all parties
(\emph{i.e.} participants) participate in a unique conversation and
are in the same state. The conversation starts with a random participant
entitled to speak - she will be called the \emph{speaker} - while
all other participants are \emph{listeners}. Other initial configurations
can of course be imposed; however, the dynamics tends towards attractors
whose basins of attraction are global (see Sec. \ref{sec:Results}):
every initial condition will thus tend to the same steady state. Hence,
for simplicity and without lack of generality, we adopt a homogeneous
initial condition. 
\item \textit{Roundtable topology.} The participants are arranged around
an ideal roundtable (\emph{i.e.} a one-dimensional torus with periodic
boundary conditions): each participant can speak with any other participant,
but she is in intimate (\emph{i.e.} spatial) contact only with her
two nearest neighbors - which define her own topological neighborhood.
This time-invariant conversational topology reproduces realistic,
spatially embedded conversations that tend to cluster in a circular-like
geometry. 
\item \textit{Turn-taking dynamics.} In a given conversation, only one person
(the speaker) speaks at any given time before another (different)
participant (a listener of the same group) is entitled to speak. Within
a single conversation, several non-overlapping sub-conversations can
nucleate. We assume for simplicity that the speakers of all sub-conversations
are appointed concurrently and simultaneously (this synchronous updating
rule can be relaxed, if needed). This rule introduces the turn-taking
dynamics in the model. 
\item \textit{Abstraction from conversational content}. We model the succession
of speakers within any given (sub-)conversational group as a stochastic
process. In principle, it is possible to use any kind of speaker-dependent
or history-dependent probability distribution to determine the choice
of the new speaker. However, in this work we wonder whether complex
patterns in the schisming dynamics can still develop without resorting
to additional and detailed individual information. This approach,
consistent with the complexity paradigm \citet{miller-page,schweitzer,Byrne},
is also coherent with content abstraction: any kind of emergent conversation
pattern will eventually appear as consequence of the cooperative behavior
mutuated by multi-party interaction, rather than of a mixture of poorly-defined
mechanisms. The probability distributions adopted in the baseline
model are uniform, \emph{i.e.} speaker- and history-independent. 
\item \textit{Joining/leaving force balance}. Participants in a specific
conversation remain in the conversation as long as they feel actively
involved in it up to their preferred degree; otherwise, they start
to wish to leave the conversation. We model this lively behavior by
assigning a degree of \textit{happiness }to each participant of the
conversation. Happiness hereby stands for \emph{e.g.} attention span,
patience, assertiveness, self-esteem, and more: it is the index of
the willingness of a participant to remain in a given conversation.
The baseline scenario has all participants initially involved in the
same table-wide conversation and assigned with the maximum level of
the happiness scale, which is set equal to that of anyone else - \emph{i.e.}
we optimistically assume a person is happy to take part in a conversation
that is about to start. Again, different initial conditions would
evolve towards the same stationary state, as we will see. The individual
happiness level is then subjected to dynamic change. It is decreased
by one unit for every conversation turn during which the participant
is not a speaker, while it is reset back to the initial level when
the person gets a new opportunity to speak. As soon as the happiness
level drops to the minimum tolerated level (set to zero in the baseline
model), the participant becomes \emph{latent, i.e.} she feels excluded
enough to watch out around her for opportunities to enter another
or a new conversation. Our parties can thus be considered as finite-states
automata with a set (ideally, a continuum) of states between the fully
conversation-integrated state (\emph{i.e.} the conversation's current
speaker or newcomers - parties with maximal happiness) and the fully-excluded
state (\emph{i.e.} the latent - parties of minimal happiness). Corollaries:
a) a speaker is always fully happy; b) a latent is necessarily a listener. 
\item \textit{Neighborhood-based schism dynamics.} When a participant is
latent, she will look to her topological neighbors to be eventually
engaged in a different conversation. She will first check whether
at least one of the neighbors is in turn latent: if this is the case,
she will start a \emph{new} conversation with her/them. This nucleation
mechanism is the responsible for the onset of schisming in our model.
She will instead join the ongoing conversation of either of her neighbors,
if such conversation differs from hers. In both cases, her happiness
level will reset to its maximum level. If none of these options are
possible, the agent remains latent, waiting for someone to talk to
her (and to return active in her previous conversation) or for someone
to go latent, or for a different conversation to take place. The use
of only local resources to escape from a stagnant conversation is
what we define as the \emph{conversational principle of least effort}. 
\end{enumerate}
To verify whether our simple assumptions capture realistic features
of real-life conversations, we implemented them and inspected the
ensuing emergent behavior in an ABM using NetLogo%
\footnote{NetLogo is available at: \href{http://ccl.northwestern.edu/netlogo/}{http://ccl.northwestern.edu/netlogo/}%
}. The simulative investigations were complemented with analytical
methods to gain further insights.

\section[Analysis and results]{The baseline model: analysis and results\label{sec:Results}}

The baseline model can be thought of as describing a homogenous group
of people leisurely engaged in chat without selection biases due to
accidental geometry, common interests, hierarchies or previous discourse
patterns.

Running the ABM with the homogeneous initial condition, it was found
that the initial table-wide conversation group splits over time into
several sub-convsersations of smaller group size. This is akin to
a spatial symmetry-breaking phenomenon: the initial, spatially-homogeneous
system (\emph{i.e.} lacking boundaries) evolves into one with spatially-defined
boundaries. This splitting process continues - despite temporary increases
of the sizes of conversation groups \textendash{} until the conversation
groups cannot split any further, that is, until each sub-conversation
reaches the absorbing state. Indeed, as long as there are more than
2 people in a (sub-)conversation, there exists always a non-null probability
that one participant will not speak before her happiness level decreases
to the minimum value, eventually driving her to leave the conversation;
this is true independent of the total number of participants in the
conversation and of their maximum happiness level. In the case of
an \emph{even} (\emph{odd}) initial number $N$ of agents, the asymptotic
configuration presents $N/2$ sub-conversations of two agents ($N/2-1$
sub-conversations of 2 agents and a single sub-conversation of three
agents). Equivalently said, the optimal though only asymptotic number
of parties in a conversation, according to the baseline model, is
predicted to be essentially $2$.

\begin{figure}[wide]
 \centering \includegraphics[width=1\textwidth]{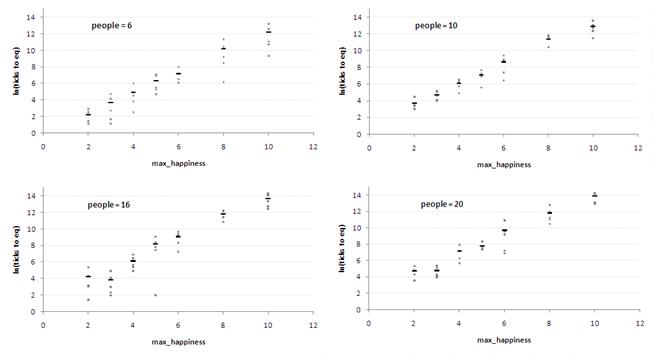} \caption{Semi-logaritmic plots of the characteristic time needed to reach the
steady state, for different settings of the maximum happiness level
(data averaged over 5 simulation runs). The straight line denotes
exponential fitting.\label{fig:Plots}}

\end{figure}

The characteristic time until reaching this steady state (\emph{i.e.}
the characteristic amount of turn-taking time steps) depends on two
factors, namely 1) the number of agents, and 2) the maximum happiness
level. As expected, if the maximum happiness level is set to infinite,
the steady state will never be reached, while if set to 1, it will
be reached very soon. Numerical simulations indicate that this characteristic
time scales exponentially with the overall maximum happiness level,
and linearly with the number of agents (Figure \ref{fig:Plots}).
The relation between happiness level and number of agents is the single
most important aspect of the model. As a matter of fact, maximum happiness
level and number of agents have opposite effects, since increase induces
an increase and a decrease, respectively, of the probability of a
single agent to be entitled as speaker. Anyway, it can be expected
that the asymptotic state is hardly reached in real-life conversations,
that typically develop within shorter timescales than the characteristic
time to stationarity. %

\subsection{Mean-field approach}

In a mean-field treatment of the baseline model, we will assume that
the probability $p_{i}$ of a participant $i$ in a conversation to
be entitled to speak by the present speaker at time t is: 1) independent
of previous conversation history, and 2) constant in time. In general,
$p_{i}=p_{i}(N,i)$ where $N$ is the number of participants, and
the specific dependence of $p_{i}$ on each participant characterizes
individuality, both intrinsic (\emph{e.g.} psychological factors)
or extrinsic (\emph{e.g.} conversation geometry). Let $F_{i}(t)$
be the happiness level of participant $i$ at time $t$; for what
said before, $F_{i}(t)$ is semipositive definite.

\subsubsection{Evolution equation}

Participant $i$ at time $t+1$ will have probability $p_{i}$ of
being a speaker - and thus of increasing $F_{i}$ to its maximum level
$MAX_{i}$, and probability $1-p_{i}$ of being a listener - thus
of decreasing her happiness level by one: $F_{i}(t)-1$. Hence we
have the following N-dimensional map $g(F_{i}(t))$: \begin{equation}
F_{i}(t+1)=p_{i}\cdot MAX_{i}+(1-p_{i})(F_{i}(t)-1),\ \forall i=1,...,N\end{equation}

\subsubsection{Fixed point and stability analysis}

To find the fixed points $F_{i}^{*}$ of each of these equations,
we drop the time dependence, \emph{i.e.}: \begin{equation}
F_{i}^{*}=p_{i}\cdot MAX_{i}+(1-p_{i})(F_{i}^{*}-1),\ \forall i=1,...,N\end{equation}
 from which we get: \begin{equation}
F_{i}^{*}=MAX_{i}+1-1/p_{i},\end{equation}
 $F_{i}^{*}$ is stable when $-1<dg(F_{i}^{*})/dF_{i}^{*}<1$. We
have: \begin{equation}
\frac{dg(F_{i}^{*})}{dF_{i}^{*}}=-p_{i},\end{equation}
 Accordingly, for normalization reasons we conclude that $F_{i}^{*}=MAX_{i}+1-1/p_{i}$
is the \emph{stable fixed point} of each participant. Now, a participant
becomes latent when $F_{i}=0$. In order for a participant to be active
in the steady state, we must have $F_{i}^{*}>0$. This translates
into $MAX_{i}>1/p_{i}-1$ which is a restriction in the waiting time
(i.e. patience) of agent $i$. Note that depending on $p_{i}$, each
agent will have a different critical patience.

As an example, in our baseline model we suppose that every agent has
the same probability of being a speaker. Imposing probability normalization,
we have $p_{i}=1/N,\ \forall i=1,...,N$. In this condition an active
steady state is achieved for $MAX_{i}>N-1,\ \forall i=1,...,N$. That
is, in order for every participant to be active in the same conversation,
their maximal waiting time cannot be less than the number of participants
minus one (the participant herself). If this requirement is fulfilled,
the initial conversation will, on average, be stable - all parties
will remain actively involved as time evolves.

\subsubsection{Extinction cascade and sociological interpretation}

The same analysis as before can be performed iteratively. Suppose
that we start at time $t=0$ with $N$ agents such that: 
\begin{itemize}
\item $p_{i}=1/N,\ \forall$ agents, 
\item $MAX_{i}>N-1\ $for $i=1,...,N-1$. 
\item $MAX_{i}<N-1$ for $i=N$. 
\end{itemize}
Then the last agent is - statistically speaking - doomed to reach
latency (and eventually leave the conversation). In order to find
the critical values of patience of the other agents, a similar analysis
as before can be performed for $N'=N-1$ and we can conclude that
the conversation will be stable if all the rest of speakers have a
patience level such that $MAX_{i}=N'-1=N-2$. Applied iteratively,
this analysis leads to the limit $N=2$ which requires $MAX_{i}>1$
- that is, a 2-party turn-taking conversation. This is consistent
with our ABM simulations.

A straightforward conclusion is the following: the number of parties
within a conversation will decrease until everybody feels comfortable
(i.e. until the patience thresholds of everybody are above the critical
values), and from there, it will remain as a stable conversation that
every speaker will profit of. The possible introduction of newcomers
into an ongoing conversation renders a direct analytic approach, even
in this very basic scenario, more difficult and goes beyond the scope
of this work. Furthermore, the analytical developments only provide
insight on the steady state, \emph{i.e.} for conversations of infinite
duration. However, as commented above, real-time conversations only
develop in finite time. Therefore, to focus on realistic scenarios
it is required to study the conversation dynamics within finite time
windows, as discussed in the following Section.

\subsection{Transient dynamics\label{sub:Transient-dynamics}}

\begin{figure}[wide]
 \centering \includegraphics[width=0.8\textwidth]{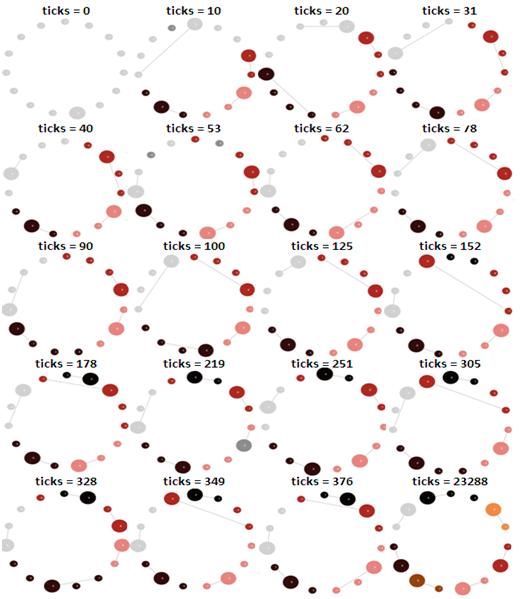} \caption{Example of transient dynamics for a 15-party conversation with maximum
happiness level of 8. Each agent is colored according to his belonging
sub-conversation; latents are colored in dark grey. The initial table-wide
conversation splits right after the beginning into 4 smaller sub-conversations
because the happiness levels of some table members necessarily become
simultaneously minimal, and latents can be mutually-close with high
probability. After the initial schism (first 10 steps), non-trivial
schisming dynamics develops, and agents hopp from a sub-conversation
to another according to the evolution of their individual happiness
status. The 4 sub- conversations persist for 150 ticks before another
sub-conversation is started. No other conversation group is formed
until 376 ticks, \emph{i.e.} the end of the table conversation.\label{fig:Example-of-transient}}

\end{figure}

The baseline model's assumption of simultaneous turn-taking (2) roughly
defines the characteristic time unit of the model (1 tick = 1 conversation
turn) as well as the empirically-relevant range of the total number
of turns taking place during a reasonable table talk. Now, what is
the correspondence between computational time steps and actual time?
While a time turn can in the real world last even very-few seconds,
here we deliberately obviate very-short time turns, since these short
turns may not have a relevant influence on the agents happiness. Accordingly,
we set a conservative estimation of an average of ten seconds per
time step (we remark that, as for the conceptual aim of the present
work, this actual value is secundary - still it will nonetheless need
to be confirmed with real experiments). A one hour-long conversation
then would allow for about 360 turns, which is hereby defined as the
\textit{actual} conversation time window. For what said before, this
information may also be used to put a lower bound on the range of
permissible maximum levels of happiness. We found that, for even and
odd numbers of participants larger than 5, avoidance of convergence
to the stationary distribution within the first 360 turns can be achieved
by setting the maximum happiness level larger than about 8 - \emph{i.e.}
8 is the minimum number of conversation turns which needs to be tolerated
without being a speaker to avoid precocious conversation convergence.
Tables of participants with higher maximum levels of happiness would
be able to maintain large conversation groups for longer periods of
time. As an example, Figure 3 shows the transient dynamics up to 376
ticks and the final stationary distribution of a model run with 15
participants and a maximum happiness level of 8. The geometric location
of, and the very participants involved in a group conversation, tend
to be persistent over time. Conversation groups rarely include people
who are not direct geometric neighbors of other people in the same
conversation. Also, latents can be trapped within a conversation group
(see \emph{e.g.} at ticks 10 and 53 in Figure \ref{fig:Example-of-transient}).
Finally, the typical size of a conversation group mildly fluctuates
in the transient timescale, assuming a typical value of $4$ parties.

\begin{figure}[wide]
 \centering \includegraphics[width=0.8\textwidth]{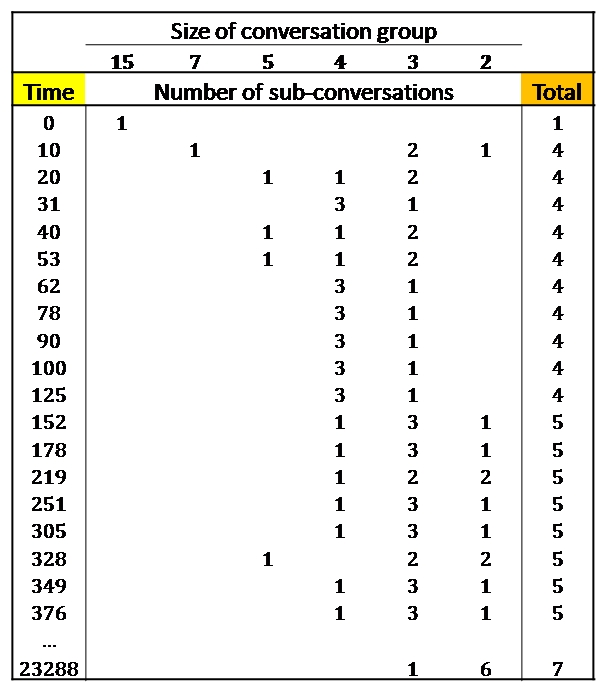} \caption{Time evolution of conversational groups for the conversation in Figure
\ref{fig:Example-of-transient}.\label{fig:Time-evolution}}

\end{figure}

All the above findings hold in general and not only in the special
illustrated example. Importantly, in spite of abstracting from conversation
contents and contexts, many of these findings seem to correspond,
at least qualitatively, to phenomena that can be observed in real
table conversations. For example: table-wide conversations involving
a large number of people are unstable, while smaller conversation
groups persist over longer periods of time; people sometimes change
conversation groups, and when this happens they confine themselves
to nearby conversations (the conversational principle of least effort
is the reason why party organizers often pay so much attention to
the initial table population and configuration, if it is supposed
to remain fixed); people within a conversation group change from time
to time, but the conversation group has a tendency to remain in a
specific geometric location, and only a limited number of people around
the table join a specific conversation group; people who have left
a conversation group may eventually return to that same conversation
later; sometimes people would like to leave a conversation, but nonetheless
they may remain in it because they are trapped between two people
eagerly taking turns in that very conversation.

The previous analysis support a fundamental conclusion: the nucleation
of sub-conversations may be considered a dynamical mechanism that
take place in conversation dynamics according to abstract and purely-local
rules of happiness evolution, also independently of context- and content-related
arguments.

\section[Empirical evidence?]{A proxy for empirical evidence?\label{sec:Proxy}}

\begin{figure}[wide]
 \centering \includegraphics[width=0.8\textwidth]{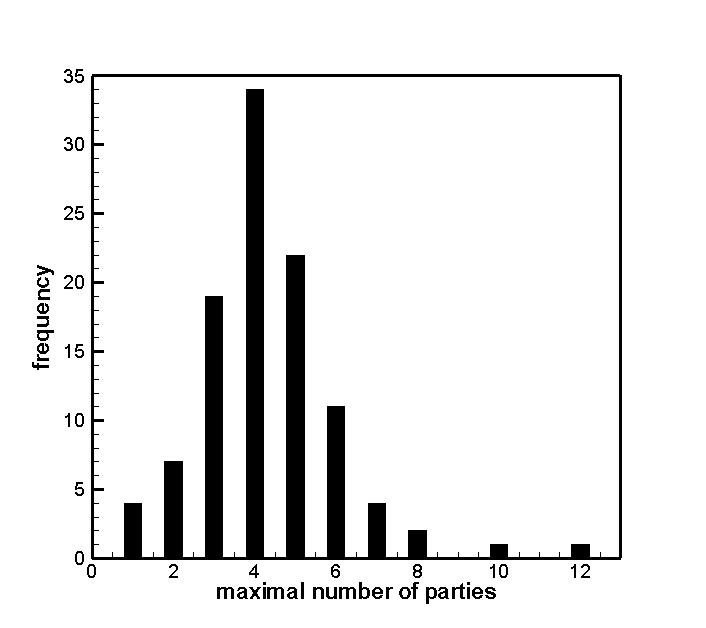} \caption{Histogram of the maximal number of people for a comfortable conversation
according to our e-mail survey. The distribution's mode ($N=4$) agrees
fairly-well with the numerical prediction of the baseline model.\label{fig:Histogram}}

\end{figure}

To assess the extent to which our model replicates quantitatively
real-world table conversation dynamics, one should compare the predicted
dynamics to large empirical data sets. While a detailed comparison
with controlled dynamical experiments is left for future investigations,
we inquired into the ABM predictions for optimal transient size of
a conversation group. We performed a preliminary e-mail poll of 105
people (with ages in the range of 20 to 40), asking the pollees to
answer to the following question: \textit{In your opinion, what is,
on average, the maximum number of people that can be involved in the
same table conversation before this conversation gets uncomfortable?}
Notice that the question does not suggest a pre-determined context-
or content-related conversation, and is free from any cognitive bias
except for the freedom implicitly allowed in the interpretation of
an uncomfortable conversation. Figure \ref{fig:Histogram} shows the
histogram of the answer's frequency. The maximal value for the size
of a stable conversation group ($N=4$) matches the typical upper
bound of conversation group sizes that were reached in our simulations
within the actual duration of conversations (see Sec. \ref{sub:Transient-dynamics})
While encouraging, this match does not suffice as empirical evidence
for the model; and we could wonder why we should assume that the opinions
of the respondents to the survey provide credible evidence. This is
statistically straightforward: people's opinion is biased on experience,
that is, their opinion is a byproduct of an average over many previous
conversations, in many different situations, scenarios, conversation
topics, conversation group characteristics, and so forth. The opinion
of individuals is therefore a proxy for real behavior. On the one
hand, such massive average over contexts and contents holds up with
our focus on context-independency; on the other hand, one could argue
that each individual is likely to have a different opinion, since
each individual is susceptible to have different experiences. Nevertheless,
if a common underlying mechanism exists, and if the histogram of individual's
opinions has a well defined average, the central limit theorem indicates
that the actual average result will tend to such average in the histogram.
Further empirical data should be obtained in order to confirm these
preliminary results.

\section[Summary]{Summary, discussion and future work\label{sec:Summary}}

The proposed, simple and abstract model of conversation dynamics predicts
a familiar behavioral scenario:

\medskip{}

I. Large conversation groups are unstable. Within a finite and reasonable
time window, we may find stable conversations of more than two people.
Schisming develops mainly from a balance between \emph{local rules}
(\emph{e.g.} happiness of parties) and the \emph{global characteristics}
(\emph{e.g.} number of participants, eventual criteria of target choice)
of the conversation.

\medskip{}

II. The formation of new conversation groups is a relatively-rare
event after the initial conversation split: the conversation dynamics
mainly consists of people joining and leaving already-existing conversation
groups, according to non-trivial spatio-temporal patterns.

\medskip{}

III. Table conversations rarely involve people who are not geometric
nearest neighbors.

\medskip{}

IV. Participants may remain trapped within their present conversation
group, in spite of their dissatisfaction.

\medskip{}

\noindent As for the (asymptotic) stationary states, we shall also
note that:

\medskip{}

V. Dyadic conversation groups are asymptotic absorbing states.

\medskip{}

VI. The characteristic time needed to reach the stationary state scales
exponentially with the maximum level of happiness, and linearly with
the number of participants.

\medskip{}

The focus of the present investigation was on the rather-conservative
baseline model of conversation dynamics, that originated as an attempt
to demonstrate a context- and content-free schisming mechanism. Further
progress in this direction will depend on the matching of simulated
and experimental data, which might well entail the refinement of the
model assumptions.

Moreover, the actual table conversation setting suggests interesting
generalizations of the model, to be tackled in further research: 
\begin{enumerate}
\item \emph{Agents heterogeneity and memory}. The baseline model has one
control parameter (the happiness degree) that can be used to fit empirical
data; also, all agents are homogenous and follow the same time-independent
behavioral rules. It seems obvious, though, that the large heterogeneity
and variety of human behavior manifests itself also in conversations.
For example, some people in a conversation group can actively try
to let people speak who have not spoken for a long time; or, on the
contrary, speakers might tend to address only people in their conversation
group who have contributed recently. As such behaviors are here modeled
by the probability distribution that determines the next speaker,
it is natural to allow for speaker-dependent and time-dependent probability
distributions, as well as for updates of the same distributions to
encode memory effects. 
\item \noindent \emph{Asynchronous updating}. Inclusion of the current speaker
in the probability function that determines the speaker of the next
turn. This eliminates the table-wide simultaneity of turn taking,
and allows a different interpretation of the characteristic time to
stationarity of the system. It also removes the stability of 2-people
conversations, and makes the stationary states potentially more interesting
- if one further assumes that 1-person conversation group cannot socially
exist, and lonely people have to join other conversation groups instead. 
\item \noindent \emph{Dynamical neighboroud topologies}: Modify the conversation
geometry so that parties can form conversation groups with more than
only two neighbors; any number of neighbors becomes possible (reminding
of \emph{e.g.} the connectivity of brain networks). A dynamic topology
might eventually reproduce cocktail party dynamics. 
\end{enumerate}
\noindent On a more abstract level, our model describes the dynamics
of interacting sub-networks where the interaction derives from random
walks taking place on these sub-networks. It would be interesting
to define fixed sub-networks and allow linkage of two different sub-networks
(\emph{i.e.} let the random walk take place on the linked sub-networks)
whenever one node in a sub-network reaches a properly-defined critical
state and joins another sub-network. These generalizations might prove
useful to model phenomena like volatility surges during financial
crises, background noise of brain activity, split and re-composition
of existing communities if regular interaction or communication is
absent, or validation frameworks for smart rooms algorithms - to cite
but a few.

\paragraph{Acknowledgments}

The authors would like to acknowledge the great hospitality and support
of the Santa Fe Institute during the Complex Systems Summer School
'09, funded by the SFI and the National Science Foundation. Special
thanks go to Dan Rockmore and Tom Carter. The authors also thank Jordi
Luque and Andrea Firrincieli for fruitful discussions. LL acknowledges
financial support from Spanish grant FIS2009-13690. Finally, the authors
would like to thank Chuck Brown for being a perpetual source of inspiration
and funk.

\newpage{}

\lhead{}
\chead{}
\rhead{REFERENCES}

\bibliographystyle{apa}
\bibliography{Roundtable}

\newpage{}

\lhead{}
\chead{}
\rhead{Figures and Captions}
\end{document}